\documentclass[aps,%
               prb,%
               reprint,%
               groupedaddress,%
               superscriptaddress,%
               longbibliography,%
               floatfix%
               ]{revtex4-2}

\usepackage[usenames,dvipsnames]{xcolor}
\usepackage{amssymb}
\usepackage{graphicx}
\usepackage{amsmath}
\usepackage{dsfont}
\usepackage{mathtools}
\usepackage{braket}
\usepackage[normalem]{ulem}

\definecolor{dark-blue}{RGB}{0,73,128}
\usepackage[colorlinks,%
            pdfusetitle,%
            urlcolor=dark-blue,%
            citecolor=dark-blue,%
            linkcolor=dark-blue]{hyperref}

\newcommand{\subfigref}[2]{\hyperref[fig:#1]{\ref*{fig:#1}(#2)}}

\newcommand{\Id}{\ensuremath{\mathds{1}}}
\DeclareMathOperator{\Tr}{Tr}
\newcommand{\D}[1]{\ensuremath{\mathrm{d}{#1}}\,}
\newcommand{\e}[1]{\ensuremath{\mathrm{e}^{#1}}\,}
\newcommand{\order}[1]{\ensuremath{\mathcal{O}(#1)}}

\renewcommand{\c}{\ensuremath{\hat{c}^{\vphantom\dagger}}}
\newcommand{\cd}{\ensuremath{\hat{c}^{\dagger}}}

\newcommand{\n}{\ensuremath{\hat{n}}}
\newcommand{\up}{\ensuremath{\uparrow}}
\newcommand{\down}{\ensuremath{\downarrow}}

\newcommand{\eps}{\epsilon}
\newcommand{\hc}{\mathrm{h.c.}}

\newcommand{\md}{\mathfrak{n}}
\newcommand{\tmd}{\md^{\mathrm{th}}}
\newcommand{\mr}{t_\down/t_\up}

\newcommand{\TUM}{\affiliation{Department of Physics, Technical University of Munich, 85748 Garching, Germany}}
\newcommand{\MCQST}{\affiliation{Munich Center for Quantum Science and Technology (MCQST), Schellingstr. 4, 80799 M{\"u}nchen, Germany}}

\begin{document}

\title{Tunable transport in the mass-imbalanced Fermi-Hubbard model}

\author{Philip Zechmann} \TUM\MCQST
\author{Alvise Bastianello} \TUM\MCQST
\author{Michael Knap} \TUM\MCQST

\begin{abstract}
The late-time dynamics of quantum many-body systems is organized in distinct dynamical universality classes, characterized by their conservation laws and thus by their emergent hydrodynamic transport.
Here, we study transport in the one-dimensional Hubbard model with different masses of the two fermionic species.
To this end, we develop a quantum Boltzmann approach valid in the limit of weak interactions.
We explore the crossover from ballistic to diffusive transport, whose timescale strongly depends on the mass ratio of the two species.
For timescales accessible with matrix product operators, we find excellent agreement between these numerically exact results and the quantum Boltzmann equation, even for intermediate interactions.
We investigate two scenarios which have been recently studied with ultracold atom experiments.
First, in the presence of a tilt, the quantum Boltzmann equation predicts that transport is significantly slowed down and becomes subdiffusive, consistent with previous studies.
Second, we study transport probed by displacing a harmonic confinement potential and find good quantitative agreement with recent experimental data~[N. Darkwah Oppong \textit{et al.}, \href{https://doi.org/10.1103/PhysRevX.12.031026}{Phys. Rev. X 12, 031026 (2022)}].
Our results demonstrate that the quantum Boltzmann equation is a useful tool to study complex non-equilibrium states in inhomogeneous potentials, as often probed with synthetic quantum systems. 
\end{abstract}
\maketitle

\section{Introduction}
Recent progress in quantum simulation and quantum computing technology enabled the realization and control of non-equilibrium quantum states of matter.
Sheer endless possibilities seem to exist to realize microscopic processes in quantum many-body systems, leading to distinct short-time dynamics.
Yet, at late times the systems' evolution coarse grains the quantum state.
In this regime, the dynamics can be grouped in a few hydrodynamic universality classes, that are solely determined by the symmetries of the system.
While the emergent hydrodynamics is generally expected to be diffusive~\cite{chaikin:2000, mukerjee:2006, lux:2014, bohrdt:2017, leviatan:2017}, recently tremendous effort has been devoted to identify quantum systems with anomalous relaxation dynamics, which can either be enhanced or suppressed.
As a consequence, different hydrodynamic universality classes have been identified.
Those range from ballistic transport in integrable models~\cite{bastianello:2022, bertini:2021} to superdiffusion in certain highly symmetric integrable models~\cite{ljubotina:2017, bulchandani:2021, wei:2022} and superdiffusion in systems with long-range interactions~\cite{schuckert:2020, joshi:2022}.
Moreover, subdiffusion can be found in systems which effectively conserve the dipole moment~\cite{gromov:2020, feldmeier:2020, guardado-sanchez:2020} and in disordered systems in the vicinity of the many-body localization transition~\cite{agarwal:2015, agarwal:2017, parameswaran:2017}.

In addition to identifying the hydrodynamic universality class, it is essential to investigate the very practical question on which timescales hydrodynamics emerges.
Extremely rich phenomenology is expected when multiple intrinsic scales are present.
In this respect, a class of systems featuring potentially interesting relaxation properties consists of a mixture of interacting particles with different single-particle masses.
For strong mass imbalance, such heavy-light mixtures have been proposed to realize a disorder-free dynamical type of many-body localization~\cite{kagan:1984, grover:2014, schiulaz:2014, schiulaz:2015}.
However, later investigations showed that these systems will relax in the thermodynamic limit, albeit on very late times, due the vastly different energy scales arising at strong mass imbalance~\cite{deroeck:2014a, papic:2015, jin:2015, yao:2016, michailidis:2018, sirker:2019, heitmann:2020a}.
Recently, this unconventionally slow relaxation dynamics of the mass-imbalanced Fermi-Hubbard model has also been experimentally observed with ultracold ytterbium atoms in optical lattices~\cite{oppong:2020}.

\begin{figure}
    \includegraphics[width=\linewidth]{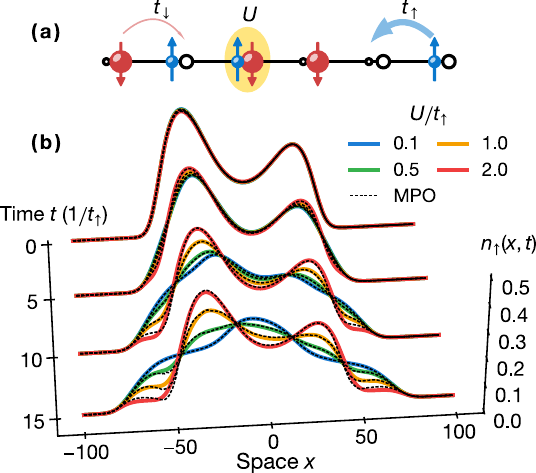}
    \caption{\label{fig:schematics_and_MPO_benchmark}
        \textbf{Dynamics in the mass-imbalanced Fermi-Hubbard model.}
        (a)~Illustration of the one-dimensional mass-imbalanced Fermi-Hubbard model, with the on-site Hubbard interaction~$U$ and distinct hopping amplitudes $t_\up$ (light) and $t_\down$ (heavy).
        (b)~Expansion after trap-release of an initial double well at inverse temperature~$\beta=0.1/t_\up$ and for fixed mass ratio~$t_\up/t_\down=0.1$.
        The kinetic theory (solid lines) is in good agreement with numerically exact calculations based on matrix product operators (dashed black line) for different interaction strengths (legend).
    }
\end{figure}

In this work, we are motivated by the question of identifying the crossover timescale to the hydrodynamic regime, by studying the mass-imbalanced Fermi-Hubbard model in the weakly interacting limit, as illustrated in Fig.~\subfigref{schematics_and_MPO_benchmark}{a}.
In this regime, the system evolves from an early-time ballistic regime to late-time diffusion, characterized solely by three conservation laws which are the energy as well as the densities of the two fermionic species.
We develop a kinetic theory based on a quantum Boltzmann equation (QBE), which is applicable to arbitrary highly excited states with no intrinsic limitations on accessible timescales.
We implement a numerical scheme to study dynamics of inhomogeneous systems for arbitrary initial states and quench protocols.
Aside from this, a linearization of the QBE directly determines the diffusion matrix by means of hydrodynamic projections~\cite{friedman:2020} and a complete characterization of the timescales that determine the crossover from ballistic to diffusive dynamics.
In particular, we show that due to the strong mass imbalance, the heavy particles strongly impede the transport of the light ones.
Despite building on the assumption of weak interactions, we show that the kinetic approach is accurate up to remarkably high interactions, by benchmarking our results with numerical tensor network simulations, which are available up to intermediate timescales [Fig.~\subfigref{schematics_and_MPO_benchmark}{b}].
For very strong interactions, deviations between exact tensor network results and QBE can be identified, indicating that multi-particle bound states can become relevant, as suggested, e.g., in Ref.~\cite{oppong:2020}, which are not captured within our kinetic theory.
\pagebreak

This work is structured as follows.
In Sec.~\ref{sec:model} we introduce the one-dimensional mass-imbalanced Fermi-Hubbard model and in Sec.~\ref{sec:boltzmann} the QBE is presented.
In Secs.~\ref{sec:linear_response} and~\ref{sec:hydro} we discuss transport in the linear response regime, and the emergence of diffusive hydrodynamics, respectively.
In particular, we find that it can take very long times for diffusion to arise, scaling as $t_\text{hydro} \sim{(\mr)}^2$, with $\mr$ the mass ratio of the two species.
External potentials can crucially modify the hydrodynamics: in Sec.~\ref{sec:tilt} we demonstrate that the QBE predicts the crossover from diffusive transport with dynamical exponent $z=2$ to subdiffusive transport with $z=4$ in the presence of a tilt potential, in line with recent experimental results~\cite{guardado-sanchez:2020} and the effective dipole-conserving hydrodynamics~\cite{gromov:2020, feldmeier:2020, burchards:2022}.
In Sec.~\ref{sec:experiment} we study the response of the system to displacing the harmonic confinement potential and find good quantitative agreement with a recent experiment~\cite{oppong:2020}.
The outlook and summary is presented in Sec.~\ref{sec:conclusion}, followed by appendices which contain the technical details.

\section{The model}\label{sec:model}
We study the one-dimensional Fermi-Hubbard model with nearest-neighbor hopping, as illustrated in Fig.\subfigref{schematics_and_MPO_benchmark}{a}, which is described by the Hamiltonian
\begin{equation}\label{eq:hamiltonian}
    \hat{H} = -\sum_{j,\sigma} t_{\sigma} \left( \cd_{j,\,\sigma} \c_{j+1,\,\sigma} + \text{h.c.} \right)
              + U \sum_{j} \n_{j,\,\up} \n_{j,\,\down}\,,
\end{equation}
where $\cd_{j,\sigma}$ ($\c_{j,\sigma}$) denotes the fermionic creation (annihilation) operator at site $j$ with spin $\sigma\in\{\up, \down\}$, $t_{\sigma}$ the species-dependent hopping amplitude, and $U$ the strength of the onsite interactions.
Here, we are interested in the case of unequal hopping matrix elements for the two spin species, and we choose the $\up$ species to be the light and the $\down$ species to be the heavy species, i.e., $t_\down/t_\up<1$.
Typically, we express all energy scales in units of $t_\up$.
The Hubbard Hamiltonian is a natural model for ultracold atoms in optical lattices, where interactions can be controlled via Feshbach resonances, while state-dependent optical lattices allow for the implementation of spin-dependent hopping amplitudes (see, e.g., Ref.~\cite{riegger:2018}).
Our model therefore describes general two-component fermionic mixtures on a lattice, where the (pseudo-)spin degree of freedom may be realized by two different nuclear spin projections or by other means.

Let us discuss the nature of transport that can be expected from general considerations.
The SU(2) symmetric one-dimensional Fermi-Hubbard model with balanced hopping, $t_\up=t_\down$, belongs to the class of integrable quantum models~\cite{essler:2005}.
Similarly, in the absence of interactions $U=0$ the model is trivially integrable for any mass imbalance.
At integrable points, transport is generically ballistic~\cite{bertini:2021,ilievski:2017, nozawa:2020, nozawa:2021}.
An important exception is the highly symmetric point with $t_\up=t_\down$ and zero total magnetization, where superdiffusion emerges~\cite{fava:2020}.
In what follows, we always assume to be in the regime where $t_\up\ne t_\down$ and interactions $U$ are small.
Therefore, we weakly depart from the trivial $U=0$ integrable point, but we do not face the complications arising from considering the highly symmetric point $t_\up=t_\down$ (see, however, Ref.~\cite{furst:2012}).
When integrability is broken, we expect on general grounds that diffusive transport of the residual conservation laws (energy and particle densities) is prevalent.
However, for significant mass imbalance the emergence of diffusion might potentially take a very long times, leading to a regime of unconventionally slow relaxation dynamics even for weak interactions.

\section{Boltzmann kinetic theory}\label{sec:boltzmann}
The quantum Boltzmann kinetic theory is a well-known approach~\cite{spohn:1991, erdos:2004}, whose derivation is based on a Bogoliubov-Born-Green-Kirkwood-Yvon (BBGKY) hierarchy of the multipoint correlation functions.
One starts with the homogeneous case and the observation that, in the absence of interactions, steady states (not necessarily thermal) are Gaussian and diagonal in momentum space.
Therefore, they are fully characterized by the two-point correlation function $W_{\sigma\tau}(k) = \braket{\cd_\sigma(k) \c_\tau(k)}$.
At the symmetric point $t_\up=t_\down$, off-diagonal terms of $W_{\sigma\tau}$ are generally non-vanishing, while the mass imbalanced case $t_\up\neq t_\down$ projects the non-trivial dynamics only on the diagonal entries.
When interactions are present, the equations of motion for $W_{\sigma\tau}$ are non-trivial and proportional to $U$, coupling to higher-order correlation functions.
While in principle all connected correlation functions are intertwined through the dynamics, the weak interactions allow for a truncation: by further invoking a separation of timescales, one can project the dynamics perturbatively on the instantaneous steady state of the non-interacting model.
Therefore, a set of closed non-linear equations is obtained for $W$: this is the quantum Boltzmann equation (QBE) $\partial_t W=U^2\mathcal{I}[W]$, where $\mathcal{I}$ is the collisional integral capturing the effects of interactions, and we factorized out the interaction dependence $\propto U^2$.
The scaling limit~\cite{furst:2013, lukkarinen:2015} formally holds in the regime of vanishing interactions and large times, in such a way that $t U^2$ is kept constant.
In practice, $1/U^2$ must be compared with the typical timescales of the non-interacting limit, set by the hopping strengths.
At this point, it is very important to distinguish the symmetric $t_\up=t_\down$ and asymmetric $t_\up\ne t_\down$ case: in the first case, the unperturbed timescale is solely determined by $t_\up^{-1}$ and one obtains a matrix-valued Boltzmann equation for $W$.
This case has been studied before~\cite{furst:2012, furst:2013, furst:2013a, lukkarinen:2015, lu:2015}, and we do not consider it in our work.
Whenever the masses are different, the unperturbed timescale is given by $\max(t_\up^{-1},t_\down^{-1},|t_\up-t_\down|^{-1})$ and the Boltzmann equation is non-trivial only on the diagonal entries $\md_\sigma(k) = W_{\sigma,\sigma}(k)$, resulting in a simplified QBE $\partial_t \md_\sigma = U^2\,\mathcal{C}_{\sigma}[\md]$.
The collision integral $\mathcal{C}_{\sigma}$ has a rather compact expression; see Appendix~\ref{sec:der_collision_op}, where we present a detailed derivation of the QBE.
For compactness, we write whenever possible $\md = (\md_\up, \md_\down)$ and $\mathcal{C} =(\mathcal{C}_\up, \mathcal{C}_\down)$ as vectors in spin space.

Weak spatial inhomogeneities can now be added to the QBE within a gradient expansion.
In this case, one assumes that the length scale of the inhomogeneity is much larger than the microscopic relaxation timescale, divided by the group velocity of the excitations.
The underlying lattice is coarse grained into a continuum variable and the mode-density is promoted to be space dependent $\md_\sigma(k)\to\md_\sigma(x, k)$.
Including the proper gradient terms, the final inhomogeneous QBE is obtained
\begin{equation}\label{eq:full_inhom_boltzmann}
    \partial_t \md_\sigma + v_\sigma(k)\,\partial_x \md_\sigma + F_\sigma\,\partial_k \md_\sigma = U^2\,\mathcal{C}_\sigma[\md]\, .
\end{equation}
Below, we also consider the addition of an inhomogeneous potential: $\hat{H}\to \hat{H}+\sum_{j,\sigma}V_\mathrm{ext}(j) \, \n_{j,\sigma}$.
As a consequence, a non-trivial force term appears with the further addition of the Hartree contribution from the interactions $F_\sigma = -\partial_x V_\mathrm{ext} - U \partial_x \int\frac{\D{q}}{2\pi} \, \md_{\bar\sigma}(x,q)$.
We denote $\bar{\sigma} =1-\sigma$ for $\up=0$ and $\down=1$, and the velocity $v_\sigma(k) = \partial_k \epsilon_\sigma(k)$ is determined by the single-particle dispersion $\epsilon_{\sigma}(k) = -2 t_\sigma\cos(k)$.

The QBE is a non-linear partial integro-differential equation, which we numerically solve by discretization of the real and momentum space, and with a mixed implicit-explicit integrator (see Appendix~\ref{sec:numerics_boltzmann}).
Energy and lattice momentum conservation (modulo $2\pi)$ fix kinematically allowed collisions, such that the collision integral only requires the numerical computation of a one-dimensional integral.
As scattering only takes place between particles of different species, incoming and outgoing particles have different dispersion, avoiding divergences in the collision integral which are present in the mass-balanced case~\cite{furst:2012, bertini:2015}.

The derivation of the QBE builds on a proper scaling limit when the interaction vanishes and inhomogeneities are smooth.
Therefore, it is of utmost importance to benchmark its validity in practical scenarios.
To this end, we compare the QBE to tensor network simulations by studying the density of the light particles $n_\up(x,t) = \int dk\ \md_\up(x,k)$ after releasing the cloud from a high-temperature thermal state in a double-well potential in Fig.~\subfigref{schematics_and_MPO_benchmark}{b}.
The benchmark with matrix product operator (MPO) simulations shows that the QBE can accurately predict the dynamics, even for comparatively strong interactions.
Note that here the operator space entanglement growth strictly limits the accessible timescales for the tensor network simulations, and the truncation error becomes significant at late times.
We would like to emphasize, that no practical limitations on the timescales and initial temperatures exist for the QBE.
%
\begin{figure*}[t]
    \includegraphics[width=\linewidth]{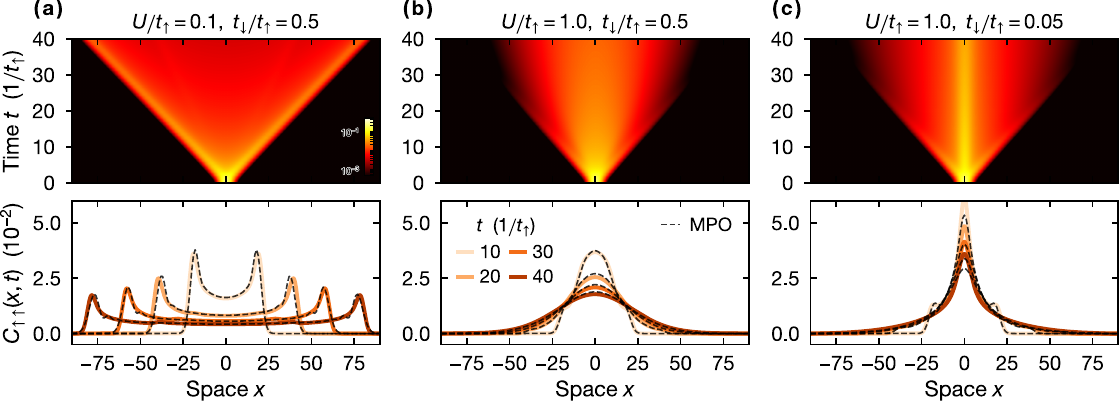}
    \caption{\label{fig:correlations_short_time}
        \textbf{Short-time evolution of correlations in the linear-response regime.}
        The density-density correlation function of the light species $C_{\up\up}(x,t)=\braket{\hat{n}_{x\up}(t) \hat{n}_{0\up}}_c$ is evaluated on a thermal background with infinite temperature at half filling and zero magnetization.
        (a)~At weak interaction $U/t_\up=0.1$ and small hopping imbalance $t_\down/t_\up=0.5$ ballistic propagation lasts for long time.
        (b)~For stronger interaction $U/t_\up=1$, correlations quickly become Gaussian, compatible with diffusive transport.
        (c)~If the mass imbalance is large $\mr=0.05$, transport is impeded by the slow species and profiles are strongly peaked and non-Gaussian on shown timescales.
        Data obtained from the QBE are compared with numerically exact matrix product operator simulations (dashed lines).
    }
\end{figure*}

\section{Transport in the linear response regime}\label{sec:linear_response}
To investigate relaxation dynamics and the crossover to diffusive transport, we study connected, unequal time density-density correlation functions of the form
\begin{equation}\label{eq:correlator}
    C_{\sigma\tau}(x,t) = \braket{\n_{x,\sigma}(t)\n_{0,\tau}}_{c} \,, 
\end{equation}
evaluated with respect to a thermal state at inverse temperature $\beta$.
This correlator is directly obtained from the kinetic theory.
It is useful to perturb the initial thermal state with an inhomogeneous chemical potential
\begin{multline}
    \beta\braket{\n_{r,\sigma}(t)\n_{0,\tau}}_{c}\\
    =\frac{\delta}{\delta \mu(0)}{\left[\frac{1}{\mathcal{Z}}\text{Tr}[e^{i \hat{H}t}\hat{n}_{\sigma}e^{-i \hat{H}t}e^{-\beta \hat{H}+\beta\sum_j \mu(j)\hat{n}_{j,\tau}}] \right]}_{\mu(j)=0}\,.
\end{multline}
Before taking the $\mu$ derivative, the above equation describes the time evolution of the density profile $\hat{n}_{r,\sigma}(t)$ evolving from an inhomogeneous initial state.
The next step is computing this object within the QBE: since in this section we are ultimately interested in linear response, we can conveniently linearize the QBE around homogeneous thermal states $\md(x,k) = \md^\mathrm{th}(k) + \delta\md(x, k)$.
By construction, thermal states are stationary solutions of the QBE $\mathcal{C}[\md^\mathrm{th}] = 0$, such that we obtain the linearization $\mathcal{C}_{\sigma}[\md] = -\sum_\tau \int\frac{\D{q}}{2\pi} \Gamma_{\sigma\tau}(k, q) \delta\md_\tau(q) + \order{\delta\md^2}$, with the linearized collision integral $\Gamma_{\sigma\tau}(k,q) = -\delta\mathcal{C}_\sigma(k) / \delta\md_\tau(q)|_{\md=\md^\mathrm{th}}$ (see Appendix~\ref{sec:der_collision_op} for the explicit expression).
On the level of the kinetic theory, the perturbation of the initial conditions due to the inhomogeneous chemical potential for computing $C_{\sigma\tau}(x,t)$ is obtained as $\delta\md_{\tau}(x, k, t=0) = \frac{1}{\beta} \partial_{\mu_\tau} \md^{\mathrm{th}}(k)\, \delta_{\tau,\sigma}\, \delta(x)$.

One has to be cautious when comparing the QBE results with lattice simulations: in contrast to the microscopic model, the kinetic equation does not have a UV cutoff.
A naive replacement of Kronecker delta with Dirac delta would induce some transient-time artifacts that, while not influencing the late-time behavior, would make a short-time comparison unfeasible.
To overcome this issue, we regularize these UV effects by broadening the Dirac delta distributions in the response functions with peaked Gaussians $\delta(x) \to \e{-x^2/2w^2} / \sqrt{2\pi}w$.
A similar coarsening procedure is then implemented on the lattice with the same width.
With that approach, the MPO simulations can be compared reasonably with the QBE and show excellent agreement.
Accuracy of the continuum approximation in particular requires a sufficiently large $w$ to eliminate oscillations resulting from the lattice, and we set $w$ to two lattice sites in all of our simulations.

At short times, the presence of interactions and mass imbalance leads to different regimes of relaxation dynamics.
In Fig.~\ref{fig:correlations_short_time} we focus on the $C_{\up\up}(x,t)$ correlator at infinite temperature.
The comparison with MPOs shows excellent agreement.
For moderate mass imbalance and interactions [Fig.~\subfigref{correlations_short_time}{a}], we find ballistic propagation with pronounced peaks at the edges of the light cone.
At later times, the system eventually becomes diffusive (not shown).
For stronger interactions [Fig.~\subfigref{correlations_short_time}{b}], the crossover to diffusion is almost immediate.
When significantly decreasing the mass ratio while keeping interactions fixed [Fig.~\subfigref{correlations_short_time}{c}], the short-time dynamics changes: the correlation profile remains narrow and is peaked for quite long times.
This is a direct consequence of the large difference in the particle masses.
The slow heavy particles strongly constrain the transport of the light particles and only at very late times transport crosses over to diffusion (not shown).
This observation demonstrates that in the limit of strong mass-imbalance relaxation dynamics can take enormously long, on timescales which can neither be accessed with exact diagonalization due to systems size limitations nor with tensor networks due to entanglement limitations.

\begin{figure}[b]
    \includegraphics[width=\linewidth]{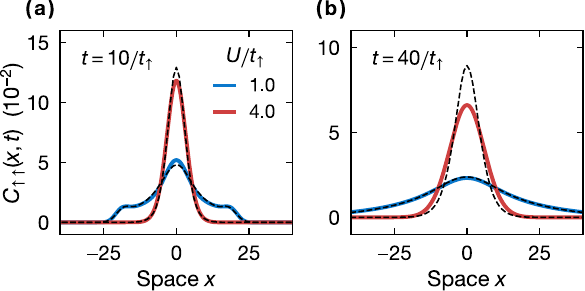}
    \caption{\label{fig:limitations}
        \textbf{Limitations of the Boltzmann theory at strong interactions.}
        We show the decay of an initially peaked correlation profile at time (a)~$t=10/t_\up$ and (b)~$t=40/t_\up$ for two interaction strengths $U/t_\up=1$ (blue) and $U/t_\up=4$ (red), where $\mr=0.2$ .
        Solid lines are obtained by solving the linearized QBE and dashed lines correspond to exact matrix product operator simulations.
        }
\end{figure}
Despite the good agreement with MPO simulations on short timescales, non-perturbative many-body effects beyond QBE are present for strong interactions [see Fig.~\ref{fig:limitations} (for $U/t_\up = 4$)].
While the profiles initially agree well, at later times the correlations decay slower in the MPO simulations than for the QBE, which we attribute to the formation of multi-particle bound states of heavy and light particles (doublons, trimers, etc.)~\cite{oppong:2020}, that are not described by the QBE.
For the remainder of this work, we focus on the regime in which the QBE is applicable and study the dynamics of the system to much later times than those accessible with the matrix product operator approach.

\begin{figure}[t]
    \includegraphics[width=\linewidth]{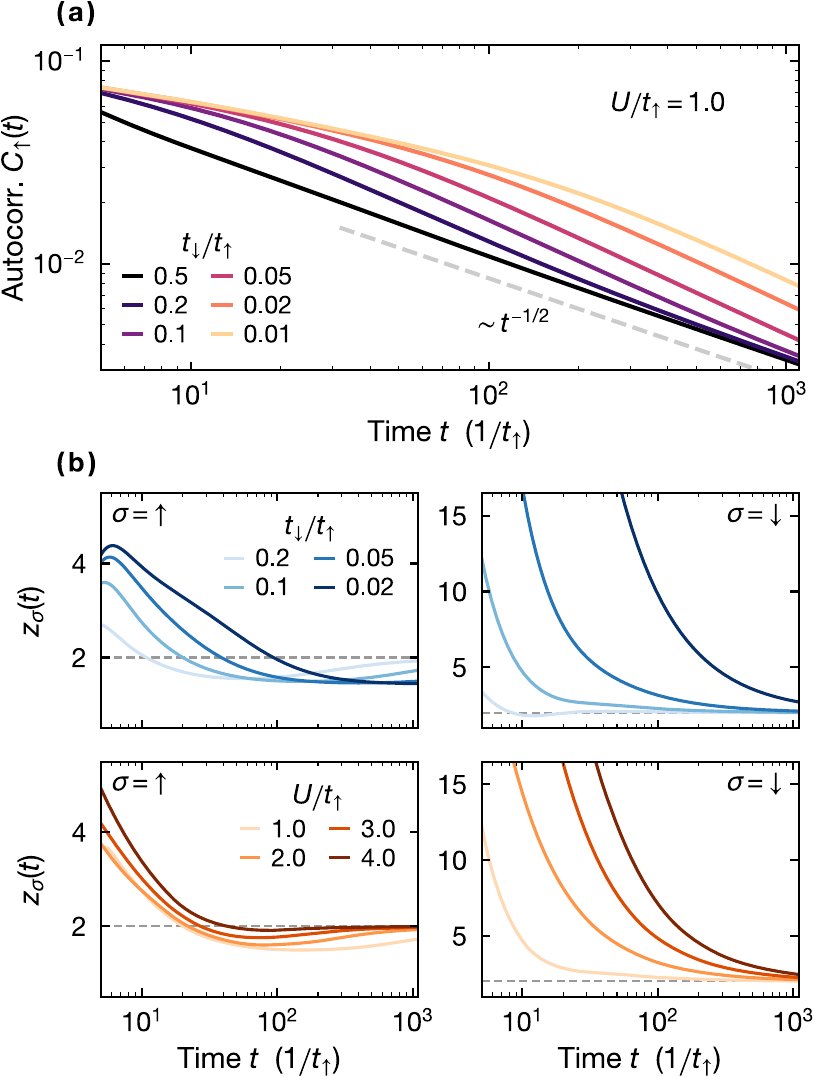}
    \caption{\label{fig:correlations_long_time}
        \textbf{Slow decay of the autocorrelation function.}
        (a)~Autocorrelation function for different mass ratios $t_\up/t_\down$ evaluated up to late times. The correlations are computed at infinite temperature for $U/t_\up=1$ with and initial Gaussian profile of with $w=2$.
    (b)~Flow of the scaling exponent of the autocorrelation function $z_{\sigma}(t) =-{[\D{\,\log C_{\sigma}} / \D{\,\log t}]}^{-1}$ for different values of $t_\up/t_\down$ at fixed interactions $U/t_\up=1$ (upper panels) and different $U$ at fixed mass imbalance $t_\up/t_\up=0.1$ (lower panels).
        The left and right columns correspond to the $\up$- and $\down$-species, respectively.
    }
\end{figure}
To study the crossover from ballistic to diffusive transport at late times, we now compute with the QBE the autocorrelation function $C_\sigma(t) = C_{\sigma\sigma}(0,t)$, see Fig.~\subfigref{correlations_long_time}{a}.
In the diffusive regime $C_{\sigma}(t) \propto 1/\sqrt{t}$.
Hence, we quantify the transient with the instantaneous dynamical exponent $-1/z_{\sigma}(t) = \D{\log C_{\sigma}} / \D{\log t}$, see Fig.~\subfigref{correlations_long_time}{b}.
Decreasing the mass ratio prolongs the transient regime for both species, and it takes very long times to reach the diffusive scaling limit with dynamical exponent $z=2$.
When increasing interactions at fixed mass ratio, light and heavy particles experience an opposite trend: while for the light particles larger interactions lead to a faster convergence of the dynamical exponent to diffusion, $z=2$, heavy particles remain very slow due to the intricate interplay of kinematics and scattering.

\begin{figure}[t]
    \includegraphics[width=\linewidth]{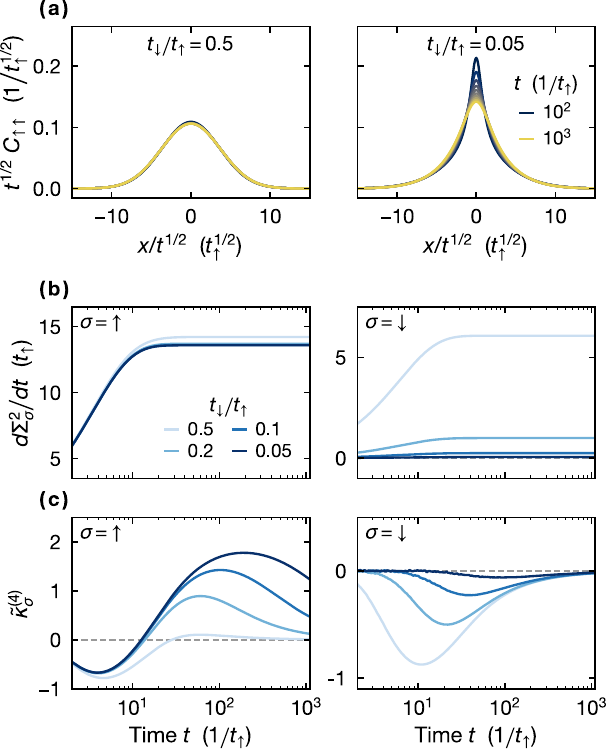}
    \caption{\label{fig:correlation_profile}
        \textbf{Correlation profiles.}
        (a)~Rescaled correlation profiles $\sqrt{t} C_{\up\up}(x / \sqrt{t}, t)$ for several times between $t \cdot t_\up=10^2\dots 10^3$, where a collapse indicates diffusive scaling.
        (b)~Spatial variance $d\Sigma^2/dt = 2D(t)$ for light (left column) and heavy (right column) particles; the late time saturation value corresponds to twice the diffusion constant.
        (c)~A finite excess kurtosis $\tilde{\kappa}^{(4)} = \kappa^{(4)}/\Sigma^2$ quantifies the non-Gaussianity of the correlation profiles.
    }
\end{figure}
To further characterize the transport of this system, we study the full correlation profiles in Fig.~\subfigref{correlation_profile}{a}.
For large mass ratio, a Gaussian profile is attained at short times.
By contrast, for a small mass ratio $\mr\ll 1$ it takes extremely long times to establish a Gaussian correlation profile, indicated by the absence of the scaling collapse.
As a consequence, for this mass ratio the transport has not reached the diffusive regime even on thousands of hopping scales.
The non-Gaussian shape of the distribution function can be understood from the extreme limit of infinitely massive heavy particles $t_\downarrow=0$.
In this limit, a light particle scattering with a heavy one can, within the quantum Boltzmann equation, at most swap the sign of the momentum as a consequence of energy-momentum conservation.
This results in a block-diagonal linearized collision integral $\Gamma_{\sigma,\sigma}(k,q)$ coupling $k$ modes only to $q=\pm k$.
As a consequence, the linearized Boltzmann equation decouples in blocks of paired momenta $(k,-k)$: each of these blocks at late time experiences diffusive behavior, with a momentum-dependent diffusion constant.
The total correlation function will thus be obtained as a weighted sum of Gaussian profiles with momentum-dependent variance, set by the $k$-dependent diffusion constants, which results in the non-Gaussian profile.
For small mass ratios, different momentum sectors couple weakly.
As a consequence, ultimately a Gaussian correlation profile will be attained, however, due to the weak coupling of the modes the non-Gaussian shape of the distribution will remain to be present for very long times; as shown in the right panel of Fig.~\subfigref{correlation_profile}{a}.

To further characterize the distribution, we study the scaling of the second and fourth cumulants of the correlation profile.
The derivative of the variance $\D{\Sigma^2}/\D{t}$ is compared for different mass ratios in Fig.~\subfigref{correlation_profile}{b}, and surprisingly we find a fast saturation of the diffusion constant $2D = \lim_{t\rightarrow\infty}\D{\Sigma^2}/\D{t}$ for all mass rations.
The non-Gaussianity of a distribution function can be characterized by the excess kurtosis, i.e., the standardized fourth cumulant $\tilde{\kappa}^{(4)} = \kappa^{(4)}/\Sigma^2$, which is zero for Gaussians.
In contrast to the width of the distribution, the excess kurtosis deviates from zero for extremely long times highlighting the non-Gaussianity of the distribution [see Fig.~\subfigref{correlation_profile}{c}].
In summary, for all considered mass ratios the scaling of the variance is compatible with diffusion on a comparatively short timescale, but the full correlations retain untypical, fat-tailed profiles for very long times.
Therefore, for large mass ratios it takes extremely long to fully establish diffusive transport for both species.
In the next section, we study these timescales in detail.

\section{The diffusive scale}\label{sec:hydro}
\begin{figure}[t]
    \includegraphics[width=\linewidth]{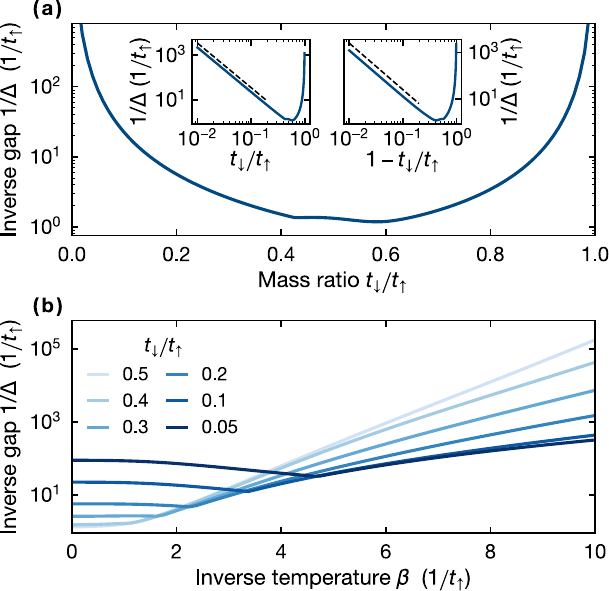}
    \caption{\label{fig:gap_scaling}
        \textbf{Scaling of the spectral gap of the linearized collision operator.}
        The gap determines the onset of diffusion at $t_\mathrm{hydro}\sim 1/\Delta$, indicating the scale when all non-conserved charges are decayed.
        (a)~Inverse gap as a function of the mass ratio at infinite temperature.
        The left (right) inset illustrates the quadratic divergence of $t_\mathrm{hydro}$ in vicinity of $t_\down/t_\up=0$ ($t_\down/t_\up$=1).
        (b)~Temperature dependence of the inverse gap $1/\Delta$ for different mass ratios.
        Note that kinks occurring in both panels correspond to crossings of the lowest eigenvalues of $\Gamma$.
    }
\end{figure}

At late time, the conservation laws of the non-interacting limit are destroyed by interactions.
Eventually, the system enters the proper diffusive limit governed by the residual conservation laws, which are the energy and the particle density of the two species.
Therefore, the late time dynamics is expected to be governed by coupled diffusive equations of the form
\begin{equation}\label{eq_diffusive_eq}
    \partial_t \delta q_\alpha - \partial_x^2 \sum_{\alpha'} D_{\alpha,\alpha'} \delta q_{\alpha'} = 0\,,
\end{equation}
where $\delta q_{1,2,3}$ denotes the local expectation values of the residual conserved charges ($\uparrow$-spin and $\downarrow$-spin particle and energy densities, respectively).
The QBE approach both determines the $3\times 3$ diffusion matrix and the diffusive timescales.
We start by studying the latter, which are readily connected with the spectrum of the linearized collision operator $\Gamma$.
Indeed, non-zero eigenvalues correspond to decaying modes, arising from collisions, and the real part of the eigenvalue is the inverse of the decay time.
Thus, the spectral gap $\Delta$ measures when the hydrodynamic regime is entered, $t_\mathrm{hydro} \sim 1/\Delta$ [Fig.~\subfigref{gap_scaling}{a}].
For intermediate mass imbalance we find $1/\Delta$ to be of $\order{1}$.
However, the gap $\Delta$ closes near $\mr=0$ and $\mr=1$, leading to divergent $t_\mathrm{hydro}$.
This is expected, as in both limits infinitely many conservation laws are present due to integrability.
In both cases we find a quadratic divergence $t_\mathrm{hydro} \sim {(\mr)}^{-2}$ and $t_\mathrm{hydro} \sim {(1-\mr)}^{-2}$, respectively, and matches our previous observation of a long ballistic-to-diffusive crossover in these regimes.
%

In Fig.~\subfigref{gap_scaling}{b} we show the temperature dependence of $1/\Delta$.
For high temperatures $\beta\lesssim 2$ the behavior is consistent with the infinite temperature case, but at low temperatures the trend is reversed: smaller mass imbalance leads to a larger $t_\mathrm{hydro}$, diverging exponentially as $\beta\to\infty$.
This can be understood from the emergence of the universal low-temperature description in terms of a Tomonaga-Luttinger liquid (with marginal perturbations)~\cite{giamarchi:2004} , which is integrable and supports ballistic transport.
At finite but small temperature, diffusion is due to excitations nearby the Fermi edges and the phase space undergoing non-trivial scattering vanishes exponentially in $\beta$.
It should be stressed that divergent timescales $t_\text{hydro}$ predicted by the QBE should be taken with care, since the Boltzmann approach captures only the first non-trivial perturbative corrections in $U$.
In proximity of these singular limits, corrections beyond QBE may become important and can modify the diffusive timescale.
Nonetheless, the divergence of $t_\text{hydro}$ within QBE is a good indicator for the extremely long times needed for diffusion to emerge.

We now evaluate the diffusion matrix $D_{\alpha,\beta}$.
Following Ref.~\cite{friedman:2020} (see also Appendix~\ref{app:hydro}), the diffusion matrix can be extracted from the QBE by integrating out the dynamics of the decaying charges and projecting on the residual conservation laws
\begin{equation}
    D_{\alpha,\alpha'} = [A \, {(P\Gamma P)}^{-1} \, A]_{\alpha,\alpha'}
\end{equation}
with the diagonal operator $A_{\sigma\tau}(k,q) = v_{\sigma}(k)\,\delta(q-k)\delta_{\sigma\tau}$ [$v_{\sigma}(k)$ is the group velocity] and $P$ a projector on the decaying modes.
Therefore, $P\Gamma P$ is invertible by construction.
%
\begin{figure}[t]
    \includegraphics[width=\linewidth]{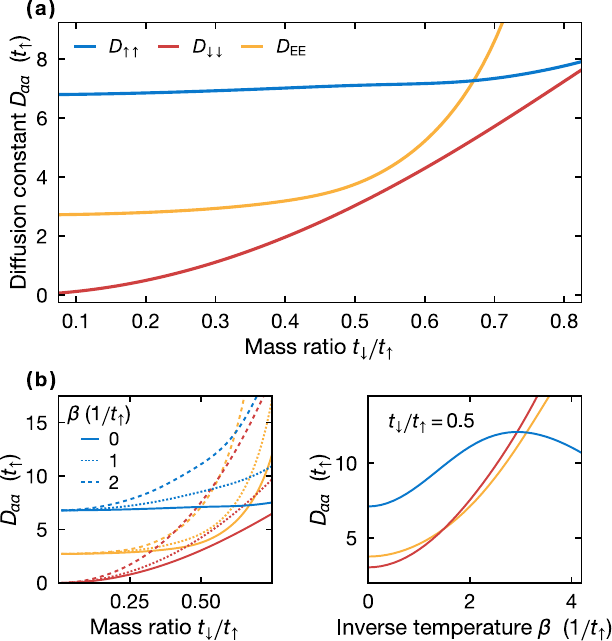}
    \caption{\label{fig:diffusion_constant}
        \textbf{Tunable diffusion constants.}
        (a)~Diffusion constants $D_{\alpha\alpha}$ as a function of  the mass ratio $t_\down/t_\up$ for $U/t_\up=1$ at infinite temperature.
        (b)~Temperature dependence of the diffusion constants.
        In the left panel the infinite temperature limit (solid) is compared to $\beta t_\uparrow=1, 2$ (dotted, dashed).
        In the right panel, the temperature dependence is plotted for $\mr=0.5$.
    }
\end{figure}
Interestingly, we find that for half-filling and zero magnetization, the diffusion matrix becomes diagonal, decoupling the hydrodynamics modes, and in Fig.~\ref{fig:diffusion_constant} we focus on this point.
With decreasing but still finite $\mr$, the $\up$ density and the energy density diffusion constants saturate to a constant value, while the diffusion constant of the $\down$ density decays as expected.
With increasing mass ratio all diffusion constants increase monotonically: as we already commented, close to mass balance the gap of the collision integral $\Delta$ closes and the diffusion matrix diverges, since at mass balance and for zero magnetization the Fermi-Hubbard model is known to exhibit superdiffusive transport~\cite{fava:2020}.

In Fig.~\subfigref{diffusion_constant}{b} we analyze the temperature dependence of the diffusion constants.
With decreasing temperature we find an increasingly stronger dependence on $\mr$.
For a fixed mass ratio the diffusion constants show a pronounced temperature dependence.
At high temperature, it follows an expansion of the form $D=D_0(1+c_2\beta^2)$~\cite{zanoci:2021}, with the infinite temperature value $D_0$, and $c_2$ a constant that depends on microscopics.
For low temperature we find a non-monotonic dependence of $D_{\up\up}$ on $\beta$, where the diffusion constant starts to decrease with decreasing temperature.

\section{Subdiffusive transport in linear potentials}\label{sec:tilt}
Generally, transport properties can be modified by external potentials. In this section we focus on a linear potential $V_\mathrm{ext}(x) = -F x$.
Non-interacting particles on a lattice in a linear potential experience Wannier-Stark localization, and perform Bloch oscillations~\cite{wannier:1960}.
Interactions can significantly affect this simple picture.
For example, a recent experimental study of a two-dimensional Fermi-Hubbard model in a linear potential~\cite{guardado-sanchez:2020} shows a crossover from diffusive transport at short wavelengths ($k\gg F/t_\up$), to subdiffusive dynamics at long wavelengths ($k\ll F/t_\up$).
In such systems with a tilted potential the coarse-grained charge dynamics is governed by an emergent hydrodynamic description equivalent to the hydrodynamics of dipole-moment conserving systems, leading to a subdiffusive mode with dynamical exponent $z=4$~\cite{gromov:2020, feldmeier:2020, guardado-sanchez:2020}.
In the limit of strong tilts, the system can even exhibit a dynamical form of localization, known as Hilbert space fragmentation~\cite{sala:2020, khemani:2020} on prethermal timescales~\cite{zhang:2020, khemani:2020, scherg:2021a, kohlert:2021}.

\begin{figure}[t]
    \includegraphics[width=\linewidth]{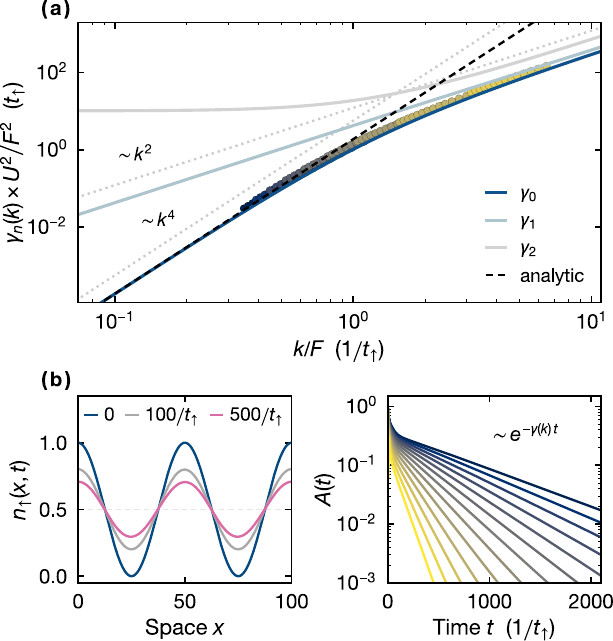}
    \caption{\label{fig:tilt}
        \textbf{Subdiffusive transport in a linear potential.}
        (a)~Decaying hydrodynamic modes $\gamma_n(k)$ obtained for $\mr=0.5$ at infinite temperature.
        The black dashed line corresponds to the analytic asymptotics, Eq.~\eqref{eq:analytic_asymptotics}, for the subdiffusive mode.
        Solid lines are extracted from linearizing the hydrodynamic equations, and symbols are obtained from solving the inhomogeneous QBE for various density wave initial conditions.
        (b)~Left panel: Decay of a density wave $n_\uparrow(x) \sim \cos(kx)$ for $\mr=0.5$, $U/t_\up=1.0$ at infinite temperature in a linear potential $V_\mathrm{ext}(x) = -F x$ at different times $t=0$ (blue), $t=100/t_\up$ (gray) and $t=500/t_\up$ (pink).
        Right panel: The amplitude $A(t)$ for a wide range of wave vectors $k$.
        The markers in (a) are obtained from the decay constants for a range of different tilts $F/t_\up\in[0.01, 0.1]$ and wave vectors $k\in[\pi/100, \pi/10]$.
    }
\end{figure}
By considering weak enough tilts and interactions strength, and thus by avoiding Hilbert space fragmentation, we can derive the subdiffusive hydrodynamics directly from our QBE.
Moreover, we can study the crossover from diffusion at short length scales to subdiffusion at long length scales.
We notice that finite-temperature homogeneous thermal states are not stationary in the presence of a tilted potential, hence, any initial state will relax to the infinite-temperature ensemble.
Therefore, we start by considering weak inhomogeneities on infinite temperature states.

As a first step we generalize the diffusive equations~\eqref{eq_diffusive_eq} to the presence of the tilted potential, which can be performed with the methods of hydrodynamic projections (see Appendix~\ref{app:hydro} for details)
\begin{equation}\label{eq:diff_tilt}
    \partial_t \delta q_\alpha-\sum_{\alpha'}{[(\partial_x-F \Sigma)D(\partial_x+F\Sigma^\dagger)]}_{\alpha,\alpha'}\,\delta q_{\alpha'}=0\,,
\end{equation}
where the matrix $\Sigma$ is defined as $\Sigma_{i,j}=\delta_{i,3}(\delta_{j,1}+\delta_{j,2})$.
The $F$-dependent shift in the diffusive equation arises because the kinetic energy $q_3$ is no longer conserved.
Instead, the total energy $e_\text{tot}(x)=q_3(x)-Fx(q_1+q_2)$ is conserved, which includes in addition to the kinetic energy also tilt contributions.
To analyze the diffusive equations~\eqref{eq:diff_tilt} we go to Fourier space and determine the eigenvalues $\gamma_n(k)$ of the operator as obtained from the zeros of $\det[\gamma_n(k)-(ik+F\Sigma)D(ik-F\Sigma^\dagger)]$.
As ${(ik+F\Sigma)}^{-1}=\frac{1}{ik}(1 -\frac{F}{ik} \Sigma)$, the eigenvalue equation can be recast in the more convenient form $\det[k^{-2}\gamma_n(k)(1-\frac{F}{ik}\Sigma)(1+\frac{F}{ik}\Sigma^\dagger)-D]=0$.
For $F/k \gg  t_\up$ to leading order $\gamma_n(k)$ solves $\det[k^{-4}F^2\gamma_n(k)\Sigma\Sigma^\dagger-D]\simeq 0$, resulting in a subdiffusive mode $\gamma_0(k)\propto k^4$
\begin{equation}\label{eq:analytic_asymptotics}
    \gamma_0(k)= \frac{k^4}{2 F^2 {[D^{-1}]}_{3,3}}\,, \hspace{2pc}k/F\ll 1/t_\up\,.
\end{equation}
In the opposite regime $k/F\gg 1/t_\up$, conventional diffusion is restored with modes $\gamma_n(k) = k^2 \lambda_n$, where $\lambda_n$ are the three eigenvalues of the diffusion matrix.
We emphasize that the crossover is solely determined by the ratio $k/F$.

The normal modes of the effective hydrodynamic equation~\eqref{eq:diff_tilt} computed from the linearized collision integral are shown in Fig.~\subfigref{tilt}{a}.
The interactions and the mass ratio are fixed far from special integrable points: we choose $U/t_\up=1.0$ and $\mr=0.5$.
In the presence of the tilt, the three diffusive modes cross-over for $k \ll F/t_\up$ to the predicted subdiffusive mode, Eq.~\eqref{eq:analytic_asymptotics}, arising from the coupling of energy and charge, a quasi-hydrodynamic mode~\cite{guardado-sanchez:2020, gromov:2020}, and a conventional diffusive mode, not present in the hydrodynamic model for a single species.
We also verified that for an $N$ species mixture, there are $N-1$ residual diffusive normal modes.
With the inhomogeneous QBE we probe typical initial states that couple to the subdiffusive mode and are employed in experimental realizations~\cite{guardado-sanchez:2020}.
As illustrated in Fig.~\subfigref{tilt}{b}, a sinusoidal perturbation for the light particles is imprinted on the initial state $n_\up(x) \sim \cos(k x)$ with a homogeneous background of heavy particles.
The amplitude $A(t)$ of the wave decays exponentially with a wave-number-dependent decay rate $\gamma(k)$.
Probing a wide range of wave vectors and tilts shows excellent agreement with Eq.~\eqref{eq:diff_tilt} for the crossover of the slowest normal mode to subdiffusion, shown as markers in Fig.~\subfigref{tilt}{a}.

\section{Far from equilibrium and experimental implications}\label{sec:experiment}
%
\begin{figure}[t]
    \includegraphics[width=\linewidth]{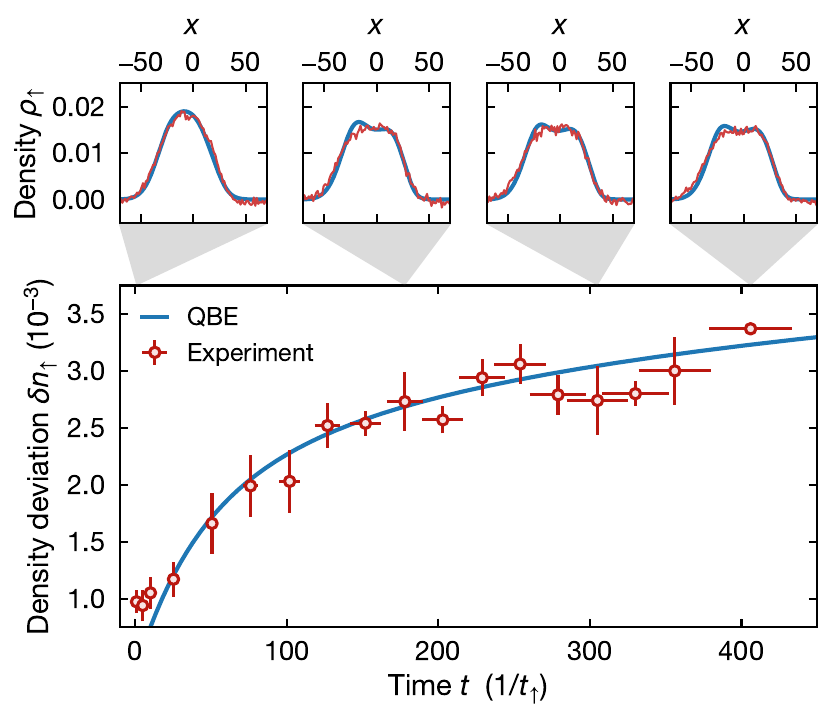}
    \caption{\label{fig:exp}
        \textbf{Comparison of the inhomogeneous QBE with experimental results of Ref.~\cite{oppong:2020}.}
        Recent experiments studied the relaxation dynamics of a heavy-light mixture of ytterbium atoms prepared in an optical lattice with harmonic confinement~\cite{oppong:2020}.
        After a slow translation of the trap minimum by $\approx20$ lattice sites, the dynamics of the light species is monitored for a time $t$.
        Experimental data are compared with the inhomogeneous QBE.
        In the upper row normalized density profiles at different points in time are shown, while the lower panel shows $\delta n_\up$, Eq.~\eqref{eq:density_dev}, which quantifies the residual dynamics.
        We show data for interaction strength $U/t_\up=-2.0$ and mass ratio $t_\down/t_\up = 0.3$.
    }
\end{figure}
The inhomogeneous QBE can be a useful tool to study the dynamics induced by involved experimental preparation schemes.
To demonstrate this, we model recent experiments on the relaxation dynamics of the mass-imbalanced Fermi-Hubbard model realized by ultracold ytterbium atoms in an optical lattice~\cite{oppong:2020}.
In this section, we use the full inhomogeneous QBE to study the non-equilibrium protocol realized in the experiment~\cite{oppong:2020}.

In the experiment, an anisotropic three-dimensional optical lattice realizes an ensemble of one-dimensional systems, which are loaded with ultracold ytterbium atoms harmonically confined by the potential $V_\mathrm{ext}=\kappa/2\,{(x-x_0)}^2$.
Mass imbalance is realized via a state-dependent optical lattice, exploiting the different polarizability of the ground state and the long-lived excited clock state of ytterbium.
On-site Hubbard interactions can be controlled with an orbital Feshbach resonance.
The system is driven out of equilibrium by displacing the trap minimum gradually over a distance of $\approx20$ lattice sites with a velocity of $\approx0.5$ lattice sites per tunneling time $1/t_\up$.
The dynamics of the light species is then monitored by \textit{in-situ} absorption imaging for different times $t$, whereby the density of the atoms $n_\up(x,t)$ is integrated over many tubes with varying atom numbers.

Studying transport in such a setup is challenging, as the preparation scheme is involved and the harmonic trap influences transport.
At the edges of the trap, in particular, the tilt can be strong enough to completely depart from a hydrodynamic approximation.
This is particularly relevant for the heavy atoms, since the local potential at the edges of the trap is large compared to the hopping $t_\down$, and transport can be significantly slowed down, or can be even in a regime of Wannier-Stark localization.
For this reason the protocol is restricted to small displacements and low fillings, where sufficiently many heavy particles are mobile.
The residual dynamics is quantified by
\begin{equation}\label{eq:density_dev}
    \delta n_\up{(t)} = {\left\{ {\int dx\,\rho_\up(x, t) {[ \rho_\up(x, t) - \rho_\up(x, 0) ]}^2} \right\}}^{1/2}\,,
\end{equation}
where the \mbox{$\rho_\uparrow(x,t) = n_\up(x, t) /\int dx\,n_\up(x, t)$} denotes the normalized particle density.
The observable $\delta n_\up{(t)}$ captures deviations from the initial state while suppressing experimental noise.

We test our QBE against the experimental realization by approximately replicating the experimental sequence numerically.
To account for the ensemble of different system sizes, we compute the weighted average of several tube sizes $N_\up=N_\down=5,\dots,30$ according to the experimentally estimated distribution~\cite{oppong:2020}.
While the experimental Hubbard parameters can be precisely estimated, the temperature of the initial state is much more challenging to characterize.
Hence, we treat it as a fitting parameter fixed by comparing the experimental density profile to our numerical initial state and obtain $T/t_\up\approx 4.5$.
In Fig.~\ref{fig:exp} we show profiles of the light species and the density deviations $\delta n_\up(t)$ over the experimentally accessible times.
The mass ratio is $\mr \approx 0.3$, and interactions are tuned to $U/t_\up \approx -2.0$.

The QBE predictions show good quantitative agreement with the experiment.
We attribute deviations, mainly visible in the central structure, to experimental noise, the finite resolution of the absorption imaging, and uncertainty in our exact knowledge of the initial state.
However, the bulk motion of the atomic cloud, quantified by $\delta n_\up(t)$, is captured remarkably well.
From this we conclude that for these parameters the system is in the kinetic regime, and the QBE faithfully describes how mass imbalance constrains the dynamics of the system.
For larger interactions $U/t_{\up}\approx -10$, which are also studied in the experiment of Ref.~\cite{oppong:2020} to demonstrate anomalously slow relaxation, our perturbative QBE is not applicable.

\section{Conclusion and Outlook}\label{sec:conclusion}
We developed a kinetic theory for the mass-imbalanced Fermi-Hubbard model in form of the quantum Boltzmann equation and studied transport using this framework.
By linearizing the quantum Boltzmann equation we computed the decay of spatio-temporal correlations within linear response and identified a very slow crossover from the ballistic to the diffusive hydrodynamic regime.
From the linearized equations we obtain the diffusion matrix and the timescale of emergent hydrodynamics, which strongly depend on the mass ratio giving rise to  anomalously slow dynamics.

Within this approach, inhomogeneous potentials can be studied as well.
Based on the Boltzmann equation, we derive the subdiffusive hydrodynamics with dynamical exponent $z=4$ for weakly tilted Hubbard chains compatible with an earlier experiment~\cite{guardado-sanchez:2020} and fracton hydrodynamics~\cite{gromov:2020, feldmeier:2020}.
Furthermore, we employ the inhomogeneous quantum Boltzmann equation to study the relaxation dynamics of a recent experimental implementation of the mass-imbalanced Fermi-Hubbard model.
We found  good agreement between the experiments and the results obtained from the quantum Boltzmann equation.
This demonstrates that the kinetic Boltzmann theory is a useful approach to study dynamics of non-equilibrium states generated by complex preparation schemes in inhomogeneous potentials, which are often realized in experiments with ultracold atoms, or other synthetic quantum systems.
Here, we focused on a one-dimensional, two-component mixture of fermions.
In principle, the formalism can be straight-forwardly generalized to higher dimensions, however, the increasing phase space for collisions leads to technical challenges.
Generalizations of the technique to multicomponent mixtures and bosonic systems are in principle straightforward and a promising route for future work.

\begin{acknowledgments}
We thank Oscar Bettermann, Immanuel Bloch, Johannes Feldmeier, Simon Fölling, Christian Mendl, Nelson Darkwah Oppong, Giulio Pasqualetti, and Brayden Ware for insightful discussions and the experimental team for providing us with the measured data.
We acknowledge support from the Deutsche Forschungsgemeinschaft (DFG, German Research Foundation) under Germany’s Excellence Strategy--EXC--2111--390814868, TRR80 and DFG grants No. KN1254/1-2, KN1254/2-1, the European Research Council (ERC) under the European Union’s Horizon 2020 research and innovation programme (Grant Agreement No. 851161), as well as the Munich Quantum Valley, which is supported by the Bavarian state government with funds from the Hightech Agenda Bayern Plus.
Matrix product operator simulations were performed using the TeNPy package~\cite{hauschild:2018}.

{\textit{Data and materials availability:}} Data analysis and simulation codes are available on Zenodo upon reasonable request~\cite{zenodo}.
\end{acknowledgments}

\bigskip
\appendix

\section{Derivation of the Boltzmann equation}\label{sec:der_collision_op}
This appendix outlines the derivation of the quantum kinetic theory for the mass-imbalanced Fermi-Hubbard model.
We consider the Fermi-Hubbard model on the infinite chain and cast the interaction terms of the Hamiltonian in a symmetric form
\begin{align} 
    \begin{split}
        \hat{H} = -&\sum_{j,\sigma} t_{\sigma} \left( \cd_{j,\,\sigma}
        \c_{j+1,\,\sigma} + \text{h.c.} \right) \\
            +&U\sum_{\mathclap{j, \{s_i\}}} I_{s_1, s_2, s_3, s_4} \, \cd_{j, s_1} \cd_{j, s_2} \c_{j, s_3} \c_{j, s_4} \,,
    \end{split}
\end{align}
by introducing the interaction vertex
\begin{equation}\label{eq:interaction_vertex}
    I_{s_1, s_2, s_3, s_4} = \frac{1}{2} \left( \delta_{s_1, s_4}\delta_{s_2,
    s_3} - \delta_{s_1, s_3}\delta_{s_2, s_4} \right) \,,
\end{equation}
which is antisymmetric under the exchange of the spin indices $s_1 \leftrightarrow s_2$ and $s_3 \leftrightarrow s_4$, and symmetric under the simultaneous exchange of the pairs $(s_1, s_2) \leftrightarrow (s_3, s_4)$.
The Hamiltonian can be translated to momentum space with the Fourier transform of the Fermi operators $\c_\sigma(k) = \sum_j \e{ikj} \c_{j, \sigma}$, which yields
\begin{equation}
    \begin{split}
    \hat{H} = \sum_{\sigma} \int
    \frac{\D{k}}{2\pi}~\epsilon_{\sigma}(k)~\cd_{\sigma}(k) \c_{\sigma}(k)
    + U\sum_{\{s_n\}} I_{s_1, s_2, s_3, s_4} \\ \times
    \int \frac{\D{k^4}}{{(2\pi)}^3}~\delta_{2\pi}(\underline{k})~\cd_{s_1}(k_1) \cd_{s_2}(k_2) \c_{s_3}(k_3) \c_{s_4}(k_4)\,,
    \end{split}
\end{equation}
where the momentum integrals are over the Brillouin zone $\mathbb{B} = [-\pi, \pi]$ with $\D{k^4} = \D{k_1} \D{k_2} \D{k_3} \D{k_4}$ and the free dispersion $\epsilon_{\sigma}(k) = -2 t_\sigma\cos(k)$.
We introduced the abbreviation $\underline{k} = k_1+ k_2 - k_3 - k_4$ for the momentum transfer, where momentum is conserved only up to $2\pi$ due to Umklapp scattering, indicated by $\delta_{2\pi}(k) = \delta(k \bmod{2\pi})$.

\subsection{Collision operator}
%
\begin{figure*}[t]
    \includegraphics[width=0.78\linewidth]{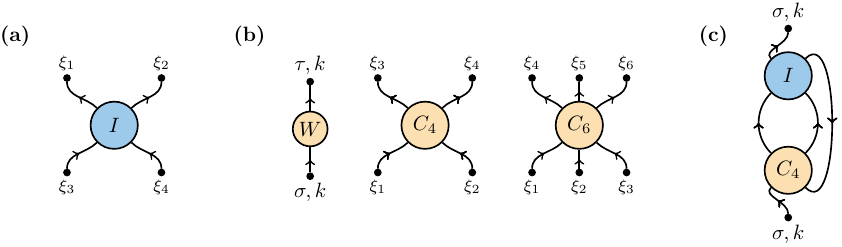}
    \caption{\label{fig:diagrams}
        \textbf{Diagrammatic notation for multipoint correlation functions.}
        (a)~The interaction tensor $I_{s_1, s_2, s_3, s_4}$ defined in Eq.~\eqref{eq:interaction_vertex}.
        (b)~The two-point $W_{\sigma\tau}$, four-point $\braket{C^{(4)}_{\{s_n\}}({k_n})}$, and six-point $\braket{C^{(6)}_{\{s_n\}}({k_n})}$ correlation functions.
        Leg indices are abbreviated by $\xi_n = (s_n, k_n)$ and the legs' positions are relevant (note the different conventions for $I$ and $C$).
        For the correlators, incoming legs connect to creation and outgoing legs to annihilation operators.
        (c)~The diagram corresponding to the right-hand side of Eq.~\eqref{eq:eom_two_point_correlator}, without the hermitian conjugate term.
        Contracting two legs amounts to integrating the momentum over $k\in\mathbb{B}$ and summing the spin index over $s\in\{\up,\down\}$.
        For each interaction tensor a global factor of $U$ must be included.
    }
\end{figure*}
For systems with weak spatial and temporal inhomogeneity, we seek to go to a kinetic description for the space-time dependent mode density describing the quasimomentum distribution, referred to as Wigner function.
The locally homogeneous system is characterized by the two-point correlation function $W_{\sigma\tau}(k) = \braket{\cd_\sigma(k)\c_\tau(k)}$, which is matrix-valued due to the presence of the two spin species~\cite{furst:2012}.
From the Heisenberg equation of motion, the evolution of the two-point correlator obeys
\begin{equation}
   i\partial_{t} W_{\sigma \sigma'}(k, t) = -\left[ \epsilon_{\sigma}(k) - \epsilon_{\sigma'}(k) \right] W_{\sigma \sigma'}(k, t) + U \times [\dots] \,.
\end{equation}
The dispersion relations of the two species are different. Hence, the off-diagonal entries of $W$ contain free contributions, which oscillate on a timescale  $t \sim |t_\up - t_\down|^{-1}\ll U^{-1}$, within the assumed scaling limit.
Consequently, on kinetic timescales off-diagonal terms in $W$ decay due to dephasing and only the diagonal correlations $\md_\sigma(k) = W_{\sigma\sigma}(k)$ matter.
For the equation of motion of the two-point correlator’s diagonal entries, we obtain
\begin{equation}\label{eq:eom_two_point_correlator}
    \begin{split}
        i\partial_t \md_\sigma(t) = U \int \frac{\D{k}^4}{{(2\pi)}^2}~\delta_{2\pi}(\underbar{k})~\delta(k_1-k) \sum_{\{s_n\}} I_{s_1, s_2, s_3, s_4} \\ \times \delta_{s_1\sigma}
        \left\{\braket{\cd_{s_1}(k_1)\cd_{s_2}(k_2)\c_{s_3}(k_3)\c_{s_4}(k_4)} + \hc \right\}\,,
    \end{split}
\end{equation}
which in turn depends on the four-point correlation function.
For Hubbard interactions the equation of motion for any $N$-point correlator will generally contain up to $(N+2)$-point correlators.
Recursively integrating these equations of motion results in a perturbative Dyson expansion in powers of $U$~\cite{furst:2013, lukkarinen:2015}.
Here, we merely outline the calculations necessary to obtain the second-order approximation.
In homogeneous settings, we can focus on the connected part of the four-point correlation function, as the Gaussian part does only give an irrelevant background contribution.

Let us abbreviate the multipoint correlators of order $N$~by~$\braket{C^{(N)}_{\{s_n\}}(\{k_n\})} \equiv \braket{C^{(N)}}$.
The equation of motion for the connected four-point correlator is then of the form
\begin{equation}\label{eq:eom_four_point_correlator}
    i\partial_t \braket{C^{(4)}}_\mathrm{c} = -\underline{\eps}\braket{C^{(4)}}_\mathrm{c} + U~\mathcal{F}{[\braket{C^{(4)}}, \braket{C^{(6)}}]}_\mathrm{c}\,,
\end{equation}
where $\underline{\eps} = \eps_{s_1}(k_1)+\eps_{s_2}(k_2)-\eps_{s_3}(k_3)-\eps_{s_4}(k_4)$ denotes the energy transfer and $\mathcal{F}$ is a functional of the multipoint correlators similar to Eq.~\eqref{eq:eom_two_point_correlator}.
Notice that in the second term we can split the correlators into their Gaussian and connected parts $U{\mathcal{F}[W]}_\mathrm{c} + \order{U^2}$, where the connected parts fulfill again a similar equation of motion leading to an additional order of $U$.
Hence, truncating these multipoint correlators at Gaussian level leaves us with an overall $\order{U^2}$ due to the factor of $U$ in Eq.~\eqref{eq:eom_two_point_correlator}, while neglecting corrections of $\order{U^3}$.
%
\begin{figure}[b]
    \includegraphics[width=0.90\linewidth]{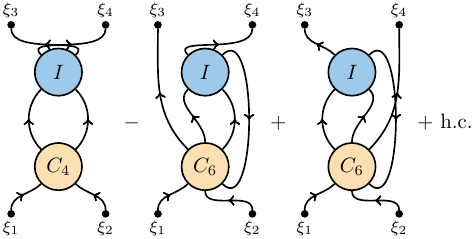}
    \caption{\label{fig:diagrams_four_point}
        \textbf{Interaction terms in the equation of motion for the four-point correlator.}
        These diagrams correspond to the functional $\mathcal{F}[\braket{C^{(4)}}, \braket{C^{(6)}}]$ in Eq.~\eqref{eq:eom_four_point_correlator}, which we want to approximate in the kinetic limit to second order in $U$.
    }
\end{figure}
%
%
\begin{figure}[b]
    \includegraphics[width=\linewidth]{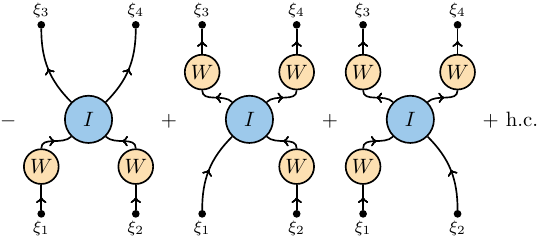}
    \caption{\label{fig:diagrams_four_point_gauss}
        \textbf{Connected part of the interaction contributions to the four-point correlator in Gaussian approximation.}
        Here, the shown diagrams represent the connected part of the functional $\mathcal{F}[W]_\mathrm{c}$.
        We obtained the diagrams in two steps:
        First, we approximate higher-order correlators by their Gaussian part, truncating the series expansion to second order in $U$;
        second, we identify all connected diagrams appearing in this approximation.
    }
\end{figure}
%
\begin{figure*}[t]
    \includegraphics[width=\linewidth]{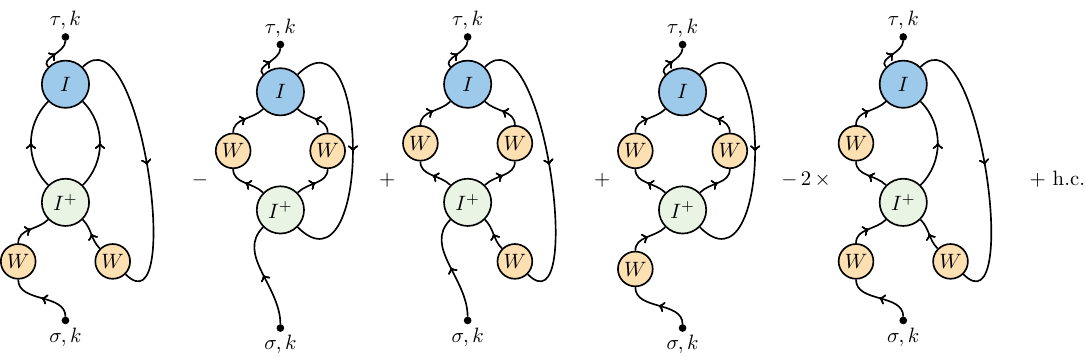}
    \caption{\label{fig:collision_diagrams}
        \textbf{Collision operator in kinetic scaling limit.}
        In Gaussian approximation, the equation for the collision operator is integrated by going to the kinetic limit, where we defined the modified interaction vertex $I^{+} = \Delta^{+} \times I$.
    }
\end{figure*}

Within this perturbative expansion, the equation of motion for the connected four-point correlator
$ i\partial_t \braket{C^{(4)}}_\mathrm{c} = -\underline{\eps}\braket{C^{(4)}}_\mathrm{c} + U~{\mathcal{F}[W]}_\mathrm{c}$
can be formally integrated~\cite{bertini:2015}
\begin{equation}
    \begin{split}
    \braket{C^{(4)}}_\mathrm{c}(t) = \, & U\int_{0}^{t}\D{t'} \e{i\underline{\eps}(t-t')t} {\mathcal{F}[W]}_\mathrm{c}  \\
    = \,& U\int_{0}^{U}\D{t'} \e{i\underline{\eps}(t-t')t} {\mathcal{F}[W]}_\mathrm{c} \\
      &+ U{\mathcal{F}[W]}_\mathrm{c} \int_{U}^{t}\D{t'} \e{i\underline{\eps}(t-t')t}\,.
    \end{split}
\end{equation}
In the second line the integral is separated into the two timescales $t<U$ and $t>U$.
In the kinetic scaling limit, we simultaneously take $U\rightarrow0$ and $t\rightarrow\infty$, while keeping $U^2t$ finite.
The first integral vanishes in this limit and for the second integral we assume the mode density to only vary slowly on the kinetic timescale, hence $\mathcal{F}$ can be pulled in front of the internal.
By regularizing the remaining integral with
$
    \int_{0}^{\infty} \D{t} \e{\pm i \omega t} = \lim_{\eta \rightarrow 0^{+}} \frac{\pm i}{\omega \pm i \eta} =%
    \pm \, i \, \mathcal{P}\left(\frac{1}{\omega} \right) + \pi \delta(\omega) = \vcentcolon\Delta^{\pm}(\omega)%
$,
we obtain for the four-point correlator
\begin{equation}\label{eq:sol_four_point}
    \braket{C^{(4)}}_{c} = -U\,\Delta^{+}(\underline{\eps})~\mathcal{F}[W]_{c} \,.
\end{equation}

As the equation of motion for the four-point correlator contains six-point correlators, computing $\mathcal{F}[W]$ in practice becomes quite cumbersome.
To this end, we make use of a diagrammatic notation to efficiently handle the bookkeeping.
In Fig.~\ref{fig:diagrams} the diagrammatic representations for the collision vertex and the multipoint correlators are shown.
The equation of motion for the two-point correlator translates to the diagram in Fig.~\subfigref{diagrams}{c}, where a factor of $U$ is associated to the interaction tensor and we need to add the hermitian conjugate.
Similarly, the diagrams corresponding to the right-hand side of Eq.~\eqref{eq:eom_four_point_correlator} are depicted in Fig.~\ref{fig:diagrams_four_point}.

At the level of six-point correlations we introduce the Gaussian approximation to truncate the Dyson expansion to second order in $U$.
Hence, we repeatedly apply Wick theorem and identify the connected contributions, depicted in Fig.~\ref{fig:diagrams_four_point_gauss}.
At this point, we keep the matrix nature of $W$, as it does not add any complications.
Discarding the off-diagonal entries later is simple and will lead to a concise expression for the collision operator.

The equation of motion for the four-point correlator can subsequently be solved in the kinetic limit, as illustrated by Eq.~\eqref{eq:sol_four_point}.
As the $W$ matrices conserve momentum, we can define $I^{\pm} = \Delta^{\pm} \times I$, and obtain the four-point correlation function from the diagrams shown in Fig.~\ref{fig:diagrams_four_point_gauss}.
Finally, plugging this result into the diagram in Fig~\subfigref{diagrams}{c} yields the expression for the collision operator, depicted in Fig.~\ref{fig:collision_diagrams}.
In principle, one could at this point convert the result back to algebraic notation.
However, in contrast to the mass-balanced case, where the collision operator can be written in a concise form in terms of products and traces of the $W$ matrix~\cite{furst:2012, furst:2013}, for unequal masses the resulting expression cannot be brought to such a form.

In the case of mass imbalance and thus assuming the dephasing of off-diagonal terms in $W$, the collision operator simplifies significantly:
\begin{equation}\label{eq:collision_operator}
    \begin{split}
        \mathcal{C}_{\sigma}[\md_{\up}, \md_{\down}] = 2U^2 \int\frac{\D{k^4}}{2\pi}~\delta_{2\pi}(\underline{k})~\delta(\underline{\eps})~\delta(k_1-k)\\
        \times \{ \md_\sigma(k_3)\md_{\bar{\sigma}}(k_4)
        [ \md_\sigma(k_1) - \md_{\bar{\sigma}}(k_2) - 1]\\
        -\md_\sigma(k_1)\md_{\bar{\sigma}}(k_2)
        [ \md_\sigma(k_3) - \md_{\bar{\sigma}}(k_4) - 1]
        \} \,.
    \end{split}
\end{equation}

Here we made use of $\Delta^{+}(\underline{\eps}) + \Delta^{-}(\underline{\eps}) = 2\pi \delta(\underline{\eps})$, hence, the principal value does not contribute, and energy is  conserved.
The resulting kinetic theory of the homogeneous model is described by the Boltzmann equation $\partial_t \md_\sigma(k) = \mathcal{C}_\sigma[\md_{\up}, \md_{\down}](k)$.
We note that, as a consistency check, it is easy to verify the number of particles and energy are exactly conserved for the stationary points describing thermal ensembles.

\subsection{Collision manifold}
The collision integral determining the kinetic description of our model is remarkably simple and only a single integral needs to be evaluated, enabling numerical studies of fully inhomogeneous settings.
Kinematically allowed collisions are defined by the collision manifold $\delta_{2\pi}(\underline{k})\delta(\underline{\eps})\delta(k_1-k)$.
For $J>0$ the set of solutions to $\{\underline{k}\bmod{2\pi} = 0 \wedge \underline{\eps} = 0\}$ has two branches.
There is the trivial solution $k_1 = k_3 \wedge k_2=k_4$, corresponding to elastic scattering and the collision integral vanishes on this contour.
Additionally, a non-trivial collision channel is present, and a closed form expression for the collision contour can be obtained.
While $k_1$ is fixed by the external momentum, we can choose to fix $k_4$ from momentum conservation and $k_3$ from energy conservation, i.e.,\ we have $\delta_{2\pi}(\underline{k}) = \delta(k_4 - f(k_1, k_2))$ and
\begin{equation}
    \delta(\underline{\eps}) = \frac{\delta(k_3 - g(k_1, k_2))}{|v_\sigma(k_3)-v_{\bar\sigma}(k_4)|}\,.
\end{equation}
Here $f(k, q) = \left[k + q - g(k, q)\right] \bmod{2\pi}$, and from some algebra we obtain for the solution
\begin{equation}
    g(k, q) = 2\arctan\left(
        \frac{\mr\sin(q+k/2) - \sin(k/2)}{\mr \cos(q+k/2) + \cos(k/2)}
    \right) \,.
\end{equation}

We note that the Jacobian can lead to singularities in the collision integral (see, for example, the the mass-balanced case~\cite{furst:2012}).
However, in our case different dispersions for $\up$ and $\down$ species avoid singular points, but nevertheless discontinuities in the integrand lead to non-analytic points of the collision integral for $\mr|\sin(k)|~\leq~1$, located at the four momenta
\begin{equation}
    k = \pm \arcsin(\mr)\,, \quad k = \pm\pi\mp\arcsin(\mr) \,.
\end{equation}
We note that each singular point is approached from one side with a square-root behavior with respect to $k$, corresponding to a divergent slope.
Such functions can be subtle for numerical integration schemes in principle. We split the integrals at these non-analytic points, but no other regularization is required.

\subsection{Linearized collision operator}
To study transport in the linear response regime we linearize the QBE around a homogeneous thermal state $\md_\sigma(k,x,t) = \md^\mathrm{th}(k) + \delta\md_\sigma(k,x,t)$, with $\md^{\mathrm{th}}_\sigma(k) = {\left[\e{\beta(\eps_\sigma(k)-\mu_\sigma)}+1\right]}^{-1} $.
By definition the collision operator vanishes for $\mathcal{C}_{\sigma}[\md_\up^\mathrm{th}, \md_\down^\mathrm{th}](k) = 0$, so we can expand to first order
\begin{equation}
    \mathcal{C}_{\sigma}(k) = -\int\frac{\D{q}}{2\pi}\sum_{\tau\in\{\up,
    \down\}} \Gamma_{\sigma,\tau}(k,q)~\delta\md_\tau(q) + \order{\delta\md^2}\,,
\end{equation}
where $\Gamma_{\sigma,\tau}(k, q) = -\left.{\delta\mathcal{C}_\sigma(k)}/{\delta\md_\tau(q)} \right|_{\md=\md^\mathrm{th}}$ denotes the linearized collision integral, obtained as variational derivative evaluated on the thermal state.
We obtain the expression
\begin{widetext}
    \begin{equation}
        \begin{split}
        \Gamma_{\sigma,\tau}(k, q) = -2U^2 \int\frac{\D{k^4}}{2\pi}~%
        & \delta_{2\pi}(\underline{k})~\delta(\underline{\eps})~\delta(k_1-k)~%
        \md^\mathrm{th}_\sigma(k_1) \md^\mathrm{th}_{\bar{\sigma}}(k_2)
        \md^\mathrm{th}_\sigma(k_3) \md^\mathrm{th}_{\bar{\sigma}}(k_4)\\
        \times \Big\{ \delta_{\sigma, \tau} & \left[
        \delta(k_1-q){[\tmd_\sigma(k_1)]}^{-2} \left(
        {[\tmd_{\bar{\sigma}}(k_2)]}^{-1} - 1 \right) -
        \delta(k_3-q){[\tmd_\sigma(k_3)]}^{-2} \left(
        {[\tmd_{\bar{\sigma}}(k_4)]}^{-1} - 1 \right) \right]\\
        + \delta_{\bar\sigma, \tau} & \left[
        \delta(k_2-q){[\tmd_{\bar\sigma}(k_2)]}^{-2} \left(
        {[\tmd_\sigma(k_1)]}^{-1} - 1 \right) -
         \delta(k_4-q){[\tmd_{\bar\sigma}(k_4)]}^{-2} \left(
        {[\tmd_\sigma(k_3)]}^{-1} - 1 \right) \right] \Big\}\,,
        \end{split}
    \end{equation}
\end{widetext}
and we simply write the action of the operator with the matrix product in spin space as $C^\mathrm{lin}(k)=-\int \frac{\D{q}}{2\pi}~\Gamma(k, q) \delta\md(q) = -(\Gamma \delta\md)(k)$.
Assuming a homogeneous background state and no external potential, the linearization of the other terms of the Boltzmann equation is straightforward, leading to the linearized Boltzmann equation
\begin{equation}
    \partial_t \delta\md_\sigma + \partial_x (A\delta\md_\sigma) - (\partial_k \md_\sigma^\mathrm{th}) \, F_\sigma
    = -\Gamma \delta\md_\sigma\,,
\end{equation}
with the diagonal operator $A_{\sigma\tau}(k,q) = v_{\sigma}(k)\,\delta(q-k)\delta_{\sigma\tau}$, and the Hartree contribution $F_\sigma = U\,\partial_x \int\frac{\D{q}}{2\pi} \delta\md_{\bar{\sigma}}(q)$.

\section{Hydrodynamic description from the method of projections}\label{app:hydro}
In this appendix, we revisit the method of projections to extract the diffusion matrix from the QBE, following Ref.~\cite{friedman:2020}.
In the first subsection, we consider the homogeneous case. The case of a tilted  potential is then discussed in the second subsection.

\subsection{Homogeneous system}
In the absence of interactions, the system has infinitely many local conserved charges in the form $q_{n}(x) = \int\frac{\D{k}}{2\pi} \langle h_{n}(k), \md(k, x) \rangle$, where $\langle h, \md\rangle = \sum_\sigma h_{\sigma} \md_\sigma$.
Once interactions are considered, the collision integral reduces the list of conserved quantities to particle number $h_{\sigma}(k) =\delta_{\sigma, \up/\down}$, and energy $h_{\sigma}(k) = \eps_\sigma(k)$.
Since we linearize close to equilibrium, we notice that the Hartree term in Eq.~\eqref{eq:full_inhom_boltzmann} can be neglected.
Furthermore, we change the basis $k\in\mathbb{B}\rightarrow {\{h_n\}}_{n=0}^{\infty}$, which yields the kinetic equation $\partial_t \delta q_n + \sum_m A_{m,n} \partial_x \delta q_m = -\sum_m \Gamma_{m,n} \delta q_m$, where both $A$ and $\Gamma$ are expressed in the charge basis.

The residual conserved charges correspond to the zero modes of $\Gamma$, where the left eigenvectors are again given by $v_{\sigma}^{1,2}(k) = \delta_{\sigma,\up/\down}$ and $v_\sigma^{3}(k) = \eps_\sigma(k)$ for particle number and energy, respectively.
Similarly, right eigenvectors are associated with thermal fixed points.
Hence, expanding $\tmd(\mu_\sigma + \delta\mu_\sigma)$ and $\tmd(\beta + \delta \beta)$ gives rise to the (unnormalized) right eigenvectors $w^{1,2}_\sigma(k) = {[1 + \cosh(\beta\eps_\sigma(k)-\mu_\sigma)]}^{-1} \delta_{\sigma, \up/\down}$ and $w^{3}_\sigma(k) = {[1 + \cosh(\beta\eps_\sigma(k)-\mu_\sigma)]}^{-1} \eps_{\sigma}(k)$.
We define $P$ the projector on the subspace of the decaying charges and $P^\perp=\Id-P$ its complement, hence $P^\perp(k) = \sum_n v^{n}(k) \otimes v^{n}(k)$.
In the charge basis, we can split the Boltzmann equation in conserved and decaying modes
\begin{subequations}\label{eq:charge_boltzmann}
     \begin{align}
         \partial_t \delta q_\alpha + \sum_m A_{\alpha, m} \partial_x \delta q_m &= 0\,, \label{eq:charge_boltzmann_line_1} \\
         \partial_t \delta q_n + \sum_m A_{n, m} \partial_x \delta q_m &= -\sum_m \Gamma_{n, m} \delta q_m\,,\label{eq:charge_boltzmann_line_2}
     \end{align}
\end{subequations}
where Greek indices correspond to conserved charges and Latin indices to decaying ones.
By inverting Eq.~\eqref{eq:charge_boltzmann_line_2} and separating out the conserved charges
\begin{equation}
    \begin{split}
        \delta q_n = -&\sum_{\alpha} {(\Gamma^{-1}\partial_t + \Gamma^{-1} A \partial_x)}_{n,\alpha}~\delta q_\alpha \\
        - &\sum_{m\neq\alpha} {(\Gamma^{-1}\partial_t + \Gamma^{-1} A \partial_x)}_{n,m}~\delta q_m\,,
    \end{split}
\end{equation}
we can iteratively express the decaying charges in terms of the conserved charges, and consider only the gradients to the lowest order~\cite{friedman:2020}
\begin{equation}
    \begin{split}
        \delta q_n =& \sum_{\alpha, m>0} {(-1)}^{m}{[{(\Gamma^{-1}\partial_t + \Gamma^{-1} A \partial_x)}^{m}]}_{n,\alpha} \delta q_\alpha\\
        \approx& -\sum_\alpha {(\Gamma^{-1} A)}_{n,\alpha} \partial_x \delta q_\alpha\,.
    \end{split}
\end{equation}
Note that matrix products are restricted to the subspace of decaying modes, such as $\sum_{l\neq\alpha} {(\Gamma^{-1})}_{n,l} A_{l,m}$.
Plugging back into Eq.~\eqref{eq:charge_boltzmann_line_1} yields the diffusion equation
\begin{equation}\label{eq_B4}
    \partial_t \delta q_\alpha - \partial_x^2 \sum_{\alpha'}{[A{(P\Gamma P)}^{-1}A]}_{\alpha,\alpha'} \delta q_{\alpha'} = 0\,,
\end{equation}
which couples the conserved modes via the $3\times3$ diffusion matrix $D_{\alpha,\alpha'} = {[A \, {(P\Gamma P)}^{-1}\, A]}_{\alpha,\alpha'}$.

\subsection{The effect of a tilted potential}
We now generalize the previous analysis to the case where a tilted potential $V(x)=-F x$ is present, deriving the hydrodynamics used in Sec.~\ref{sec:tilt}.
When a tilted potential is activated, infinite-temperature states are the only homogeneous steady state of the Boltzmann equation.
Therefore, we linearize charge fluctuations around this state.
We can straightforwardly repeat the same procedure as before, with the caveat that Eqs.~\eqref{eq:charge_boltzmann} now account for the presence of the external force and thus become
\begin{subequations}
     \begin{align}
         \partial_t \delta q_\alpha + \sum_m A_{\alpha, m} \partial_x \delta q_m +F\sum_{m} B_{\alpha,m}q_m &= 0\,,  \\
         \partial_t \delta q_n + \sum_m A_{n, m} \partial_x \delta q_m +F\sum_{m} B_{n,m}q_m &= -\sum_m
         \Gamma_{n, m} \delta q_m\,.
     \end{align}
\end{subequations}
The $B_{j,j'}$ operator originates from the gradient of the potential in Eq.~\eqref{eq:full_inhom_boltzmann}.
In the momentum basis, we have $B_{\sigma\tau}(k,q)=\delta_{\sigma\tau}\,\delta(k-q)\,\partial_q$.
By repeating the same analysis as before, but considering also the deformation of the equations induced by the weak potential, one obtains the modified diffusion equation
\begin{equation}\label{eq_B6}
    \partial_t \delta q_\alpha - \sum_{\alpha'} {[(A\partial_x+F B){(P\Gamma P)}^{-1}(A\partial_x+FB)]}_{\alpha,\alpha'} \delta q_{\alpha'} = 0\,,
\end{equation}
which we now further simplify.
While $A$ and $B$ are different operators, a simple relation can be established in the basis of the charges. In particular, $B_{1,j}=B_{2,j}=0$ holds for every $j$ and for the energy index a simple relation with $A$ holds $B_{3,j}=-(A_{1,j}+A_{2,j})$. This immediately follows from comparing the matrix elements
\begin{multline}\label{eq_B_A}
    B_{3,j}=\sum_\sigma\int\frac{\D{k}}{2\pi}\epsilon_\sigma(k)\partial_k h_j(k)\\
    =-\sum_\sigma\int\frac{\D{k}}{2\pi}\partial_k\epsilon_\sigma(k) h_j(k)=-(A_{1,j}+A_{2,j})\,,
\end{multline}
where above we integrate by parts and used the definition of the group velocity $v_\sigma(k)=\partial_k \epsilon_\sigma(k)$.

Using Eq.~\eqref{eq_B_A}, together with the fact that $B$ is antisymmetric, we can replace $P^\perp BP= -\Sigma P^\perp A P$ and $P BP^\perp=PAP^\perp \Sigma^\dagger$, where $\Sigma_{\alpha,\alpha'}=\delta_{\alpha,3}(\delta_{\alpha',1}+\delta_{\alpha',2})$.
Using this identity in Eq.~\eqref{eq_B6} and the diffusion constant implicitly defined in~\eqref{eq_B4}, finally yields the hydrodynamic equation~\eqref{eq:diff_tilt}.

\section{Numerical methods}\label{sec:numerics_boltzmann}
The simple structure of the collision integral allows for a numerical solution of the inhomogeneous non-linear kinetic theory, described by Eq.~\eqref{eq:full_inhom_boltzmann}.
For this purpose we discretize the partial differential equation in real space and momentum space on a uniform grid ${\left\{k_n = -\pi + l\Delta k\right\}}_{l=0}^{N_k} \times {\left\{x_m = m\Delta x\right\}}_{m=0}^{N_x}$, with spacing $\Delta k = {2\pi}/{N_k}$ and $\Delta x = {L}/{N_x}$, such that ${(\md_\sigma)}_{l, m} = \md_\sigma(k_l, x_m)$, and we simplify the notation again by $\md = {(\md_\up,\md_\down)}^T$.
Hence, at each step in time the mode density is approximated by an $N_k\times N_x$ real matrix for each spin species.
Such a discretization, known as method of lines~\cite{schiesser:1991}, reduces our problem to an ordinary integro-differential equation, where space and momentum derivatives are approximated by finite differences.
We use the second-order central discretization $\partial_x \md_{l,m} = [\md_{l,m+1}-\md_{l,m-1}]/2\Delta x+ \order{\Delta x^2}$, and similar for the momentum derivative.
For both real and momentum space we impose periodic boundary conditions.
In principle, other boundary conditions may be used in real space without any technical complications.

Our Boltzmann equation is in the form of a continuity equation for the mode density describing convective motion in phase space.
As expected for such a problem, we find poor numerical stability with explicit solvers.
The use of fully implicit schemes, such as the commonly employed Cranck-Nicolson method~\cite{crank:1947}, requires a prohibitively large number of evaluations of the collision integral, as the implicit equation must be solved at every time step.
For this reason, we use a mixed implicit-explicit method~\cite{ascher:2006}, where the collision integral is treated explicitly and the convective terms implicitly, such that we can separate
\begin{equation}
    \partial_t \md = f[\md] + g[\md]\,,
\end{equation}
with $f[\md]=-v\,\partial_x\md - F\,\partial_k\md$ and $g[\md] = U^2\, \mathcal{C}[\md]$. Specifically, we discretize the time domain $\md^{n} = \md(t_n)=\md(n\,\Delta t)$ for some appropriate time step, and use the Crank-Nicolson-Adams-Bashforth scheme~\cite{ascher:2006}
\begin{equation}
    \begin{split}
        \frac{1}{\Delta t} \left( \md^{n+1} - \md^{n} \right )
        =\,&\frac{1}{2} \left( f[\md^{n+1}] + f[\md^{n}] \right) \\
        +&\frac{3}{2} g[\md^n] - \frac{1}{2} g[\md^{n-1}] \,,
    \end{split}
\end{equation}
which is exact to second order in the time step $\Delta t$.
It applies Crank-Nicolson to the implicit part and the two-stage Adams-Bashforth to the explicit part.
The algebraic equation in the implicit step can be solved for $\md^{n+1}$ via fixed point iterations.
Note that $f[\md]$ is a non-linear functional in $\md$ due to the self-consistent dependence of the Hartree term on $\md$, and furthermore, that root-finding with the Newton-Raphson algorithm would require computing the Jacobian of size $(2N_{x}N_k)\times(2N_{x}N_k)$, which is both expensive and memory consuming.
Hence, we define the map
\begin{equation}
    \Phi[\md] = \md^{n} + \frac{\Delta t}{2} \left( f[\md] +
    f[\md^{n}] + 3\,g[\md^n] - g[\md^{n-1}] \right) \,,
\end{equation}
for which the new value is a fixed point $\Phi[\md^{n+1}] = \md^{n+1}$, and iterate $\md^{n+1}_{(j+1)} = \Phi[\md^{n+1}_{(j)}]$ until convergence is reached for some given accuracy threshold $||\md^{n+1}_{(j+1)}-\md^{n+1}_{(j)}|| < \delta_\mathrm{FP}$.
For the initial value a forward Euler step $\md^{n+1}_{(0)} = \md^n + \Delta t \left( f[\md^{n}] + g[\md^{n}] \right)$ is used as a first estimate.

At each time step, the collision integral needs to be evaluated once on the space-momentum grid, and we use standard Gauss-Legendre quadrature to accurately compute it.
Note that for $J|\sin(k)|\leq 1$ the integrand has up to two discontinuities, and in this case we split up the integration domain appropriately.
Our algorithm proves stable for a wide range of parameters and external potential.
We typically use $N_k, N_k \sim 200$ and $\Delta t = 0.02-0.1$ and $\delta_\mathrm{FP}=10^{-12}$.

\section{Tensor network simulations}
By means of tensor network simulations we calculate dynamical correlation functions of the form $\braket{\hat{q}_{j}\hat{q}_{j'}(t)}$, evaluated on a background equilibrium state $\braket{\hat{O}} = \Tr[\hat{O}\,\hat{\rho}]$, where $\hat{q}_j$ is the local density of a conserved charge.
For simplicity, we focus on infinite temperature $\hat{\rho} = \Id^{\otimes L}/\mathcal{N}$, with $L$ the system size and $\mathcal{N}=4^L$ the Hilbert space dimension.
We can efficiently represent the initially local operator $\hat{q}_j$ as a matrix product operator (MPO) in form of a product operator only acting non-trivial at the $j^\mathrm{th}$ lattice site.
The unitary time evolution in the Heisenberg picture $i\partial_t \hat{q}_j = i[\hat{H}, \hat{q}_j]$ can be computed with standard tensor network methods.
For this purpose, the MPO is represented as a matrix product state (MPS) in a doubled Hilbert space by combining the physical legs.
For such a vectorized operator $\ket{\hat{q}_j}$ the time evolution $\partial_t \ket{\hat{q}_j}= \mathcal{L} \ket{\hat{q}_j}$ is governed by the Liouvillian superoperator $\mathcal{L}=\hat{H}\otimes\Id-\Id\otimes\hat{H}$, and is solved by $\ket{\hat{q}_j}=\e{i\mathcal{L}t} \ket{\hat{q}_j}$.
We calculate the time evolution by Trotterization of the time evolution superoperator with the well-established time-evolving block decimation (TEBD) algorithm~\cite{vidal:2004, zwolak:2004}.
Our implementation is based on the TeNPy package~\cite{hauschild:2018}.
Due to translational invariance it is sufficient to carry out the computation once for the central site $\ket{\hat{q}_{L/2}(t)}$ and subsequently obtain the full correlation profile at each time step by applying $\hat{q}_j$ and computing the trace.

Similarly, the time evolution of the system under a quench $\hat{H} \rightarrow \hat{H}'$ can be computed.
For this purpose, we obtain the density matrix at finite temperature $\hat{\rho}=\e{-\beta\Hat{H}}/\mathcal{Z}$, which subsequently can be evolved under a quenched Hamiltonian $\hat{H}'$.
We use the vectorization as a purification of the density matrix $\ket{\hat{\rho}}$ to obtain an MPS representation.
Starting from the maximally mixed infinite-temperature state $\hat{\rho} = \Id^{\otimes L}/\mathcal{N}$ imaginary-time evolution up to $\beta/2$ yields the thermal state, and the von Neumann equation is solved by real-time evolution with the Liouvillian $\ket{\hat{q}_j}=\e{-i\mathcal{L}'t} \ket{\hat{q}_j}$.
Both calculations are efficiently carried out with TEBD, and we can thereafter evaluate observables, such as the particle density.
Generally, the maximal evolution time is limited by the growing operator-space entanglement, where we fix the maximal bond dimension to $\chi_\mathrm{max}=512$.

\bibliography{references}

\end{document}